\def\targetformat{arXiv}

\ifx\targetformat\undefined	
\def\clsstyle{prl} 

\newcommand{\appref}[1]{the Supplemental Materials}
\newcommand{\arxivtext}[1]{}
\newcommand{\prltext}[1]{#1}
\else		
\def\clsstyle{pra} 

\newcommand{\appref}[1]{App.~\ref{#1}}
\newcommand{\arxivtext}[1]{#1}
\newcommand{\prltext}[1]{}

\fi

\documentclass[aps,\clsstyle,reprint,twocolumn,showpacs,superscriptaddress,footnoteinbib]{revtex4-1}
\usepackage{graphicx,amssymb,amsmath,psfrag,xcolor,bm,xfrac}
\usepackage{bbm}
\usepackage[colorlinks=true,citecolor=blue,linkcolor=blue,urlcolor=blue]{hyperref}
\usepackage{braket}
\usepackage[colorinlistoftodos]{todonotes}

\newcommand{\n}{\hat{n}}

\usepackage{framed}
\definecolor{shadecolor}{gray}{0.95}

\usepackage{amsthm}
\newtheorem*{theorem*}{Theorem}
\begin{document}
	\title{Quotient symmetry protected topological phenomena}
	
\author{Ruben Verresen}
\address{Department of Physics, Harvard University, Cambridge, MA 02138, USA}
\author{Julian Bibo}
\address{Department of Physics, Technical University of Munich, 85748 Garching, Germany}
\author{Frank Pollmann}
\address{Department of Physics, Technical University of Munich, 85748 Garching, Germany}
\affiliation{Munich Center for Quantum Science and Technology (MQCST), D-80799 Munich, Germany}

\begin{abstract}
Topological phenomena are commonly studied in phases of matter which are separated from a trivial phase by an unavoidable quantum phase transition. This can be overly restrictive, leaving out scenarios of practical relevance---similar to the distinction between liquid water and vapor. Indeed, we show that topological phenomena can be stable over a large part of parameter space even when the bulk is strictly speaking in a trivial phase of matter. In particular, we focus on symmetry-protected topological phases which can be trivialized by \emph{extending} the symmetry group. The topological Haldane phase in spin chains serves as a paradigmatic example where the $SO(3)$ symmetry is extended to $SU(2)$ by tuning away from the Mott limit. Although the Haldane phase is then adiabatically connected to a product state, we show that characteristic phenomena---edge modes, entanglement degeneracies and bulk phase transitions---remain parametrically stable. This stability is due to a separation of energy scales, characterized by quantized invariants which are well-defined when a subgroup of the symmetry only acts on high-energy degrees of freedom. The low-energy symmetry group is a quotient group whose emergent anomalies stabilize edge modes and unnecessary criticality, which can occur in any dimension.
\end{abstract}

\maketitle

\ifx\targetformat\undefined
\textbf{Introduction.}
\else
\section*{Introduction}
\fi
In the presence of symmetries, ground states of many-body quantum systems can form \textit{symmetry protected topological} (SPT) phases of matter, which cannot be diagnosed by conventional local order parameters.
Characteristic for SPT phases is their anomalous symmetry action at the edge \cite{Pollmann10,Turner11,Fidkowski11,Schuch11,Chen11,Chen13,Wen13,Kapustin14a,Kapustin14b} implying gapless edge modes \cite{Kennedy90}, degenerate or gapless entanglement spectra \cite{AKLT88,Li08,Pollmann10}, and nonlocal order parameters \cite{dennijs89,Perez08,Haegeman12,Pollmann12}.

A celebrated one-dimensional (1D) example is the Haldane phase of the spin-$1$ antiferromagnetic Heisenberg chain, protected by, e.g., $SO(3)$ spin rotation symmetry \cite{Haldane83a,Haldane83b,AKLT88}. The edge transforms as a spin-$1/2$, which---due to its fractional nature---cannot be gapped out while preserving $SO(3)$ symmetry \cite{Pollmann12b}. The same phase arises in a bond-alternating spin-$1/2$ Heisenberg chain, $H = \sum_j (1+(-1)^j \delta) \bm S_{j} \bm{\cdot S}_{j+1}$. No gapped spin-rotation symmetric Hamiltonian can adiabatically connect the two $\delta<0$ and $\delta>0$ regimes. Within the \emph{two-site} unit cell, the spin rotation symmetry forms an integer $SO(3)$ representation. If lattice sites start at $j=1$, then $\delta>0$ is the topological phase, with a zero-energy spin-$1/2$ at each edge. This Hamiltonian adiabatically connects to the spin-$1$ Heisenberg chain as $\delta \to +\infty$ \cite{Hida92}.

\begin{figure}[ht]
    \includegraphics[width=\columnwidth]{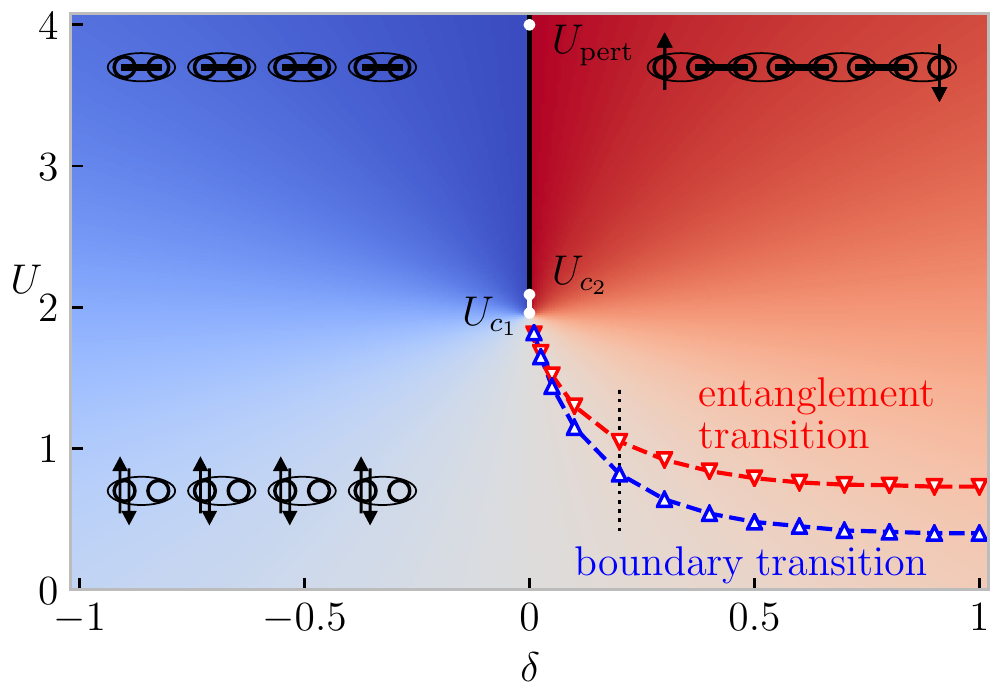}\
	\caption{\textbf{Emergent Haldane SPT phase in ionic Hubbard chain with $\Delta=0.4$.} The Mott limit $U \to \infty$ recovers two topologically distinct spin chains which are connected by the small-$U$ regime. The solid black (white) line is a second (first) order phase transition.	The transition line ends in Ising criticality at $U_{c_1}$. 
	The edge mode of the (emergent) Haldane phase is stable until it gaps out at small $U$ (blue dashed line) without a bulk phase transition. The red dashed line indicates where degeneracy of lowest entanglement level is lost.
	}
	\label{fig:Fig1}
\end{figure}

In recent times, it has been realized that the property that makes SPT phases non-trivial---the unusual symmetry action at its boundary---is its Achilles' heel. Any SPT phase can be trivialized by extending its symmetry group~\cite{Wang2018,Prakash2018,Tachikawa2020,Thorngren2020,Prakash20}. For instance, the Haldane phase is trivialized by extending $SO(3)$ into $SU(2)$. This corresponds to introducing degrees of freedom on which the $-1$ element of $SU(2)$ acts as an operator, rather than a number (such that it remains $SU(2)$ even after blocking sites into unit cells). This can be achieved by driving the spin chain away from its Mott limit, such that $2\pi$-rotations equal the fermion parity operator, rather than a classical number. Refs.~\cite{Anfuso07,Moudgalya15} demonstrated that using a fermionic model, one can adiabatically connect the trivial and Haldane phase. In this work, we ask: \textit{how immediate is this trivialization?} Are topological characteristics---such as edge modes---immediately gapped out by some exponentially-small energy scale?
Or do the edge modes remain \emph{exact} zero-energy edge modes over some finite region of parameter space?

We show that---remarkably---the latter option is realized. When an SPT phase is, strictly speaking, trivialized by introducing new degrees of freedom on which the symmetry group acts in an extended way, we argue that while certain features such as string orders are known to immediately lose their meaning \cite{Anfuso07}, other salient features such as edge modes, degeneracies of the low-lying entanglement spectrum and bulk phase transitions are stable over a finite region in parameter space. We identify quantized invariants which characterize this stability. Hence, the original symmetry group---which is now a quotient group of the extended symmetry group---can still protect certain topological phenomena; we refer to these as \emph{quotient symmetry protected topological (QSPT)} phenomena.

We illustrate this parametric stability by considering the ionic Hubbard chain \cite{Nagaosa1986,Fabrizio99,Torio01,Brune02,Kampf03,Manmana04,Batista04,Otsuka05,Torio06,Garg06,Craco08,Tincani09,Go11,Hafez14} with bond-alternation.
In the Mott limit, this reduces to the bond-alternating Heisenberg chain. As shown in Fig.~\ref{fig:Fig1}, these two SPT phases can be adiabatically connected by tuning away from the Mott limit. This phase diagram also shows two key novel features: (i) the edge modes remain stable until one drives a boundary phase transition, and (ii) the SPT criticality is stable over a finite region. After explaining these phenomena, we put them in the more general context of emergent anomalies, showcasing their applicability to general dimensions and symmetry groups.

\ifx\targetformat\undefined
\textbf{Bond-alternating Ionic Hubbard model (BIHM).} We consider spinful fermions governed by
\else
\section*{Bond-alternating \\ Ionic Hubbard model}
The bond-alternating ionic Hubbard model (BIHM) of spinful fermions consists of three terms: 
\fi
$\hat{H} = \hat{H}_\delta + \hat{H}_\Delta + \hat{H}_U$ with
\begin{align}
\hat{H}_\delta &= -\sum_{j,s}\left[\left(1+(-1)^j \delta \right)\hat{c}^{\dagger}_{j+1,s}\hat{c}^{\vphantom{\dagger}}_{j,s} +\textrm{h.c.} \right]\notag\\
\hat{H}_\Delta &= \frac{\Delta}{2}\sum_{j,s}(-1)^{j}\hat{n}_{j,s} \label{eq:ham}\\
\hat{H}_U &= U\sum_{j}\left(\n_{j,\uparrow}-\frac{1}{2}\right)\left(\n_{j,\downarrow}-\frac{1}{2}\right). \notag
\end{align}
Whenever this Hamiltonian has a unique ground state, it is naturally at half-filling \footnote{Particle-hole transformation combined with bond-centered inversion is a symmetry.}; however, this is not essential for what we discuss.

The symmetry of interest is $SU(2)$ spin-rotation generated by $\hat{S}^{\gamma=x,y,z} =\frac{1}{2}  \sum_{s,s'} \hat c_{n,s}^\dagger \sigma^\gamma_{s,s'} \hat c_{n,s'}^{\vphantom \dagger}$. A key aspect---which we will often return to---is that $2\pi$ rotations $e^{2\pi i \hat S^\gamma}$ equal the fermion parity \emph{operator} $\hat P$ rather than a classical \emph{number} $\pm 1$ as in spin chains. Hence, even after blocking into unit cells, the symmetry remains $SU(2)$ instead of becoming $SO(3)$.
%
The $SU(2)$ symmetry cannot stabilize a non-trivial SPT phase, i.e., it does not admit any non-trivial projective representations, or in terms of group cohomology \footnote{Strictly speaking, one should consider graded projective representations, but the same conclusion holds. Moreover, the fermionic nature of the underlying particles is in fact circumstantial: similar conclusions would hold for spinful hardcore bosons, albeit less physically motivated.}: $H^2(SU(2),U(1)) = 0$. In passing we mention that everything we discuss carries through if we only consider the discrete quaternion subgroup $\mathbb Q_8 \subset SU(2)$ generated \footnote{Note that, e.g., $\hat R_x^2 = \hat R_y^2 = \hat P$ and $\hat R_x \hat R_y = \hat P \hat R_y \hat R_x$. More generally, this is a representation of the quaternion group: $\rho(i) = \hat R_x$, $\rho(j) = \hat R_y$, and $\rho(-1)= \hat P$.} by the $\pi$-rotations $\hat R_x$ and $\hat R_y$, which also cannot protect the Haldane SPT phase.

In the Mott limit $U \to \infty$, the ground state has exactly one particle per site. The system is then described by an effective spin-$1/2$ chain whose Hamiltonian at leading order in $1/U$ is given by the bond-alternating Heisenberg chain. Fermion parity becomes a classical number: $P=-1$ per site, or $P=1$ for a two-site unit cell. The symmetry group per unit cell is the quotient group \footnote{If $H$ is a normal subgroup of $G$, the quotient group $G/H$ is defined by identifying every element of $H$ with the identity element.}  $SO(3) = SU(2)/\mathbb Z_2^f$ which is well-known to protect a non-trivial SPT phase, i.e., $H^2(SO(3),U(1)) = \mathbb Z_2$. Indeed, in the introduction we saw how $\delta>0$ ($\delta<0$) is a topological (trivial) phase. Similarly, $\mathbb Z_2 \times \mathbb Z_2 = \mathbb Q_8 / \mathbb Z_2^f$ also protects the Haldane phase.

Group-theoretically, the Haldane phase is thus trivialized by including fluctuating charge degrees of freedom. This has been explored before \cite{Anfuso07,Moudgalya15} and is a particular instance of how extending a symmetry group (here, extending $SO(3)$ by fermion parity into $SU(2)$) can trivialize any SPT phase \cite{Wang2018,Prakash2018,Tachikawa2020,Thorngren2020,Prakash20}.
The question of interest explored in this work is to which extent the topological features immediately disappear, i.e., are they fine-tuned or not? As a concrete model for exploring this question, we now show how the ionic Hubbard chain gives a gapped symmetric path from the Haldane phase to the trivial phase. Note that the ionicity $\Delta \neq 0$ in Eq.~\eqref{eq:ham} is necessary for explicitly breaking bond-centered inversion \cite{Pollmann12b}, $SO(4)$ and anti-unitary particle-hole symmetry \cite{Verresen17} which would otherwise still distinguish two SPT phases. Using the density matrix renormalization group \cite{White92,White93,Hauschild18}, we obtain the phase diagram with ionicity $\Delta=0.4$, shown in Fig.~\ref{fig:Fig1}.

This model has been studied before at $\delta=0$ \cite{Nagaosa1986,Fabrizio99,Torio01,Brune02,Kampf03,Manmana04,Batista04,Otsuka05,Torio06,Garg06,Craco08,Tincani09,Go11,Hafez14}. In the $U \to + \infty$ limit, it reduces to the gapless spin-$1/2$ Heisenberg chain. Although translation symmetry is still absent (due to $\Delta \neq 0$), the effective spin chain is translation-invariant at all orders in $U$~\cite{Nagaosa1986}. Remarkably, the criticality persists far beyond this perturbative regime. To highlight this, we denote the point $(U_{\rm pert})$ where perturbation theory for the translation invariant Hubbard model ($\delta=\Delta=0$) in $1/U$ diverges~\cite{Essler2005}. Along the line $\delta=0$ the model shows two quantum phase transitions: first, a BKT-transition $(U_{c_2})$ into a spontaneously dimerized insulator (SDI) phase and afterwards, an Ising transition $(U_{c_1})$ into a bond insulator (BI) phase. Whereas previous studies of this model have been field-theoretic or numerical, we will explain the persistence of this critical line as a being a quotient symmetry protected topological phenomenon. We will identify a quantized lattice invariant which implies this stability, readily extending to other symmetry groups.

For $\delta \neq 0$, we observe that this model adiabatically connects the Haldane and trivial SPT phase by passing through this fermionic regime. This is similar to the findings of Anfuso and Rosch in Ref.~\cite{Anfuso07} based on a Hubbard ladder. In this work we study the stability of the edge modes; we will see that these are stable until encountering a boundary phase transition (blue dashed line in Fig.~\ref{fig:Fig1}). A similar behavior is observed for the entanglement spectrum. Finally, note that these boundary and entanglement transitions merge at $U_{c_1}$, where the critical line disappears, suggesting a universal reason for these features.

\begin{figure}
    \includegraphics[width=0.45\textwidth]{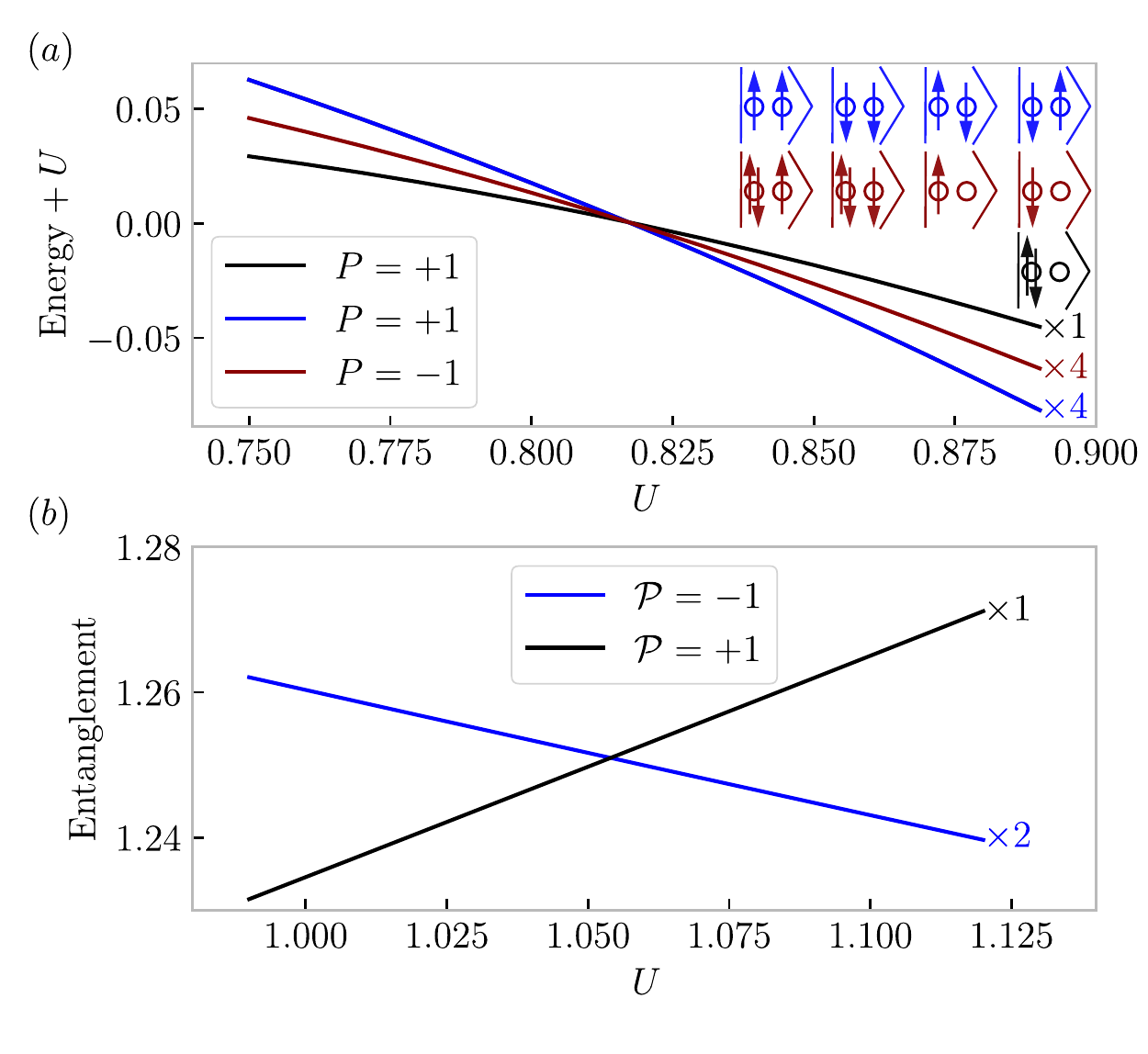}
	\caption{\textbf{QSPT edge phenomena.} We evaluate parity quantum numbers for a vertical slice $\delta=0.2$ (dotted line in Fig.~\ref{fig:Fig1}).
	(a) The ground state degeneracy and parity quantum numbers ($P$) for a system with open boundaries. We observe a boundary phase transition where the parity gap closes below (above) which the ground state is unique (degenerate).
	(b) The three lowest entanglement levels and the quantum numbers
	of $\hat{\mathcal P} = \hat{\mathcal U}_x \hat{\mathcal U}_y \hat{\mathcal U}_x^\dagger \hat{\mathcal U}_y^\dagger$ where $\hat{\mathcal U}_\gamma$ is the action of $\hat R_\gamma$ on the space of dominant eigenstates of the reduced density matrix for a bipartition of an infinitely-long chain \cite{Pollmann10}.
	}
	\label{fig:2}
    \end{figure}

\ifx\targetformat\undefined
\textbf{Stability of edge modes.}
\else
\section*{Stability of edge modes}
\fi
Here we show that the zero-energy edge modes of the Haldane phase are stable until the fermion parity gap closes for open boundary conditions. Since gaps are parametrically stable, this implies that the topological edge modes exist over a finite region of parameter space, i.e., they do not vanish as soon as $U$ is finite.

Per the usual arguments of symmetry fractionalization, an on-site unitary symmetry $\hat U$ acts on a symmetric gapped chain with edges as $ \hat U = \hat U^L \hat U^R$ \cite{Fidkowski11,Turner11}. Here, $\hat U^{L,R}$ are exponentially localized on the boundaries. As discussed above, a fermionic chain with $SU(2)$ symmetry cannot host a non-trivial SPT phase; hence, these fractionalized symmetries must obey the same group properties as the bulk symmetry, in particular: $\hat R_x^L \hat R_y^L = \hat P^L \hat R_y^L \hat R_x^L$. In the Mott limit $U\to \infty$ for $\delta >0$, we know the edge is a spin-$1/2$ degree of freedom, i.e., $P^L = -1$ such that $\hat R_x^L \hat R_y^L = - \hat R_y^L \hat R_x^L$ implies a twofold degeneracy. An eigenvalue of $\hat P^L$ cannot immediately jump such that its associated twofold degeneracy is parametrically stable. The only way $P^L$ can change is if we make $U$ sufficiently small such that another (non-degenerate) level with $P^L = +1$ crosses it in energy.

As an illustration, it is instructive to consider $\delta = 1$, where the edge mode decouples from the bulk. Focusing on the left edge (i.e., site $j=1$ in Eq.~\eqref{eq:ham}), we read off that the energy of a doubly-occupied site (where $P^L = +1$) is $\frac{U}{4}-\Delta$ whereas the energy of a singly-occupied site (with $P^L = -1$) is $-\frac{U}{4}-\frac{\Delta}{2}$. Hence, in this simple case, the edge mode is stable until $U = \Delta$, at which point it undergoes a boundary phase transition (which in $0+1D$ corresponds to a level crossing).

Away from this exactly-solvable limit, we numerically determine this boundary transition as shown in Fig.~\ref{fig:Fig1}. In the many-body system, we cannot directly read off the eigenvalue of $\hat P^L$, instead we focus on $\hat P =\hat P^L \hat P^R$. The ground state always satisfies $P=+1$ (since both edges happen to undergo the boundary transition at the same time); nevertheless, the above reasoning shows that the gap for this global parity does close with open boundary conditions (since it must close at each edge) which we verify in Fig.~\ref{fig:2}(a). The above arguments can be repeated for a virtual bipartition of an infinite system, with the degenerate low-lying entanglement spectrum being stabilized by the parity quantum numbers of the dominant eigenstates of the reduced density matrix, confirmed in Fig.~\ref{fig:2}(b).

\ifx\targetformat\undefined
\textbf{Stability of SPT transition.}
\else
\section*{Stability of SPT transition}
\fi
Here we explain why the phase transition between a trivial and SPT phase does not immediately gap out after extending the symmetry group.
Consequently, the critical line in Fig.~\ref{fig:Fig1} is a generic QSPT phenomenon.

Our model has a useful duality symmetry $\delta \to - \delta$ given by the modified (unitary) translation operator $\hat{\mathcal D}\hat c_{n,s} \hat{\mathcal D}^\dagger = (-1)^n \hat c_{n+1,s}^\dagger$.
A direct transition can thus only occur at $\delta=0$.
In the spin chain limit, $\hat{\mathcal D}$ acts as a single-site translation symmetry and an on-site unitary: $\hat{\mathcal D} \hat{S}^\gamma_j \hat{\mathcal D}^\dagger = \hat R_y \hat{S}^\gamma_{j+1} \hat R_y^\dagger$. It is well-known that in combination with spin-rotation symmetry, this implies a Lieb-Schultz-Mattis (LSM) anomaly, disallowing a gapped symmetric ground state \cite{Lieb61,Chen11_LSM,Cheng16,Cho17,Yang18,Metlitski18,Jian18}: in the absence of symmetry breaking, this stabilizes a direct phase transition. The standard proofs for the LSM anomaly hinge on the fact that on a single site, $\hat R_x$ and $\hat R_y$ anticommute. This no longer holds when charges fluctuate: $\hat R_x \hat R_y \hat R_x^{-1} \hat R_y^{-1} = \hat P$ (and indeed, in Fig.~\ref{fig:Fig1} we see that small $U$ admits a gapped phase). Nevertheless, we show that there is an \emph{emergent Lieb-Schultz-Mattis theorem}, enforcing the parametric stability of the phase transition.

Above, we understood the parametric stability of edge modes in terms of parity quantum numbers. Similarly, we now introduce a quantized invariant for a parity string. As long as fermionic operators remain gapped, the fermion parity string generically has long-range order. Moreover, at $\delta=0$, this must have a well-defined momentum under the duality/translation symmetry $\mathcal D$:
\begin{equation}
    \langle \hat P_{m} \hat P_{m+1} \cdots \hat P_{n-1} \hat P_n \rangle \sim \textrm{constant} \times  e^{i\theta(n-m)}. \label{eq:Pstring}
\end{equation}
Since parity is a $\mathbb Z_2$ symmetry, we obtain a quantized invariant $\theta \in \{ 0, \pi \}$. Eq.~\eqref{eq:Pstring} can be rigorously derived from symmetry fractionalization (see Appendix~\ref{app:theta}). Intuitively, $\theta=\pi$ formalizes the idea of being close to a Mott limit, where fermion parity coincides with the parity of the number of sites.

\begin{figure}
    \includegraphics[width=0.45\textwidth]{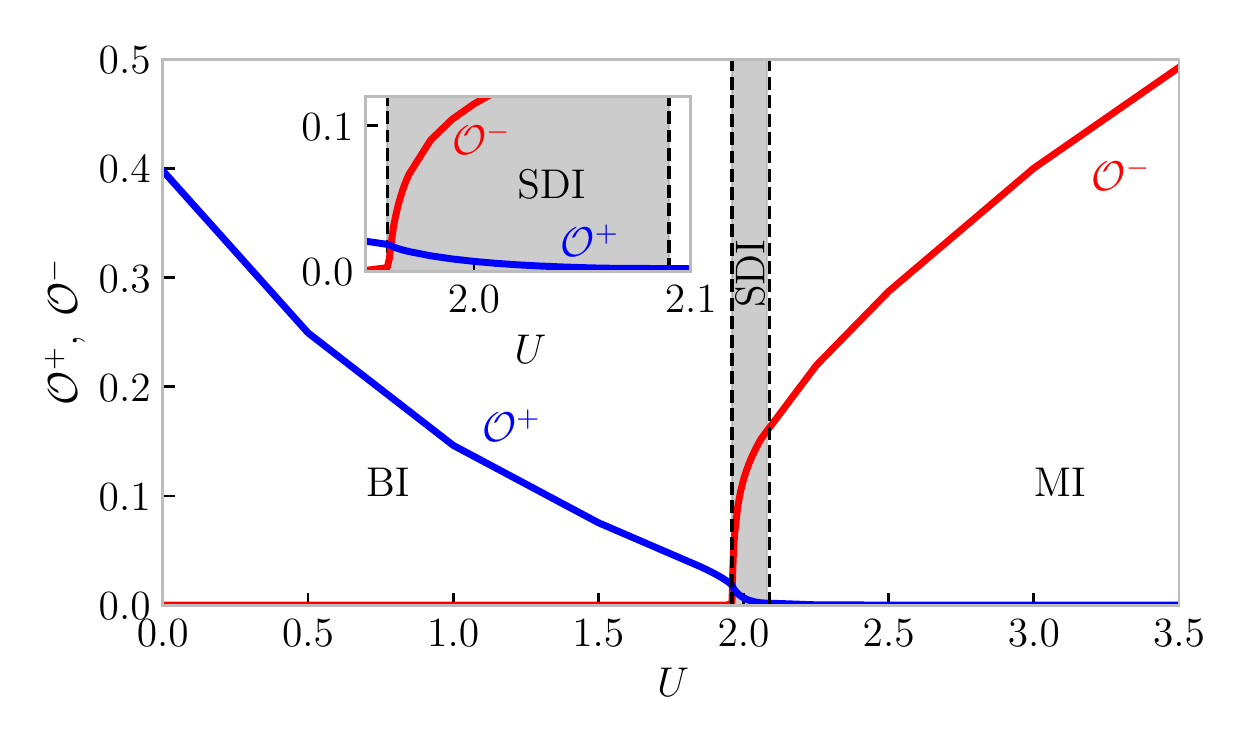}
    \caption{\textbf{QSPT transition and emergent anomaly along the self-dual line $\delta = 0$ (with ionicity $\Delta =0.4$).} The observable $\mathcal O^\pm$ captures whether the fermion parity string oscillates or not: it measures the $\mathbb Z_2$ invariant $\theta=0,\pi$ in Eq.~\eqref{eq:Pstring}. If $\theta=\pi$ (i.e., $\mathcal O^- \neq 0$ and $\mathcal O^+ = 0$), there is an emergent Lieb-Schultz-Mattis anomaly, preventing a symmetric gapped state. Inset: the intermediate phase spontaneously breaks the duality symmetry $\mathcal D$ such that both $\mathcal O^\pm$ are nonzero. \label{fig:3}}
\end{figure}

We claim that if $\theta = \pi$, then there is an emergent anomaly that forbids a gapped symmetric ground state. To prove this, suppose there is such a ground state. Symmetry fractionalization then allows us to factorize a string of $\hat R^x$ as $\hat R^x_L \hat R^y_R$. Now consider the composite symmetry $\hat{\mathcal I} = \hat R_y \hat{\mathcal D} \hat R_y^\dagger \hat{\mathcal D}^\dagger$.
Since $\mathcal D$ and $R_y$ commute, $\hat{\mathcal I} = 1$. Alternatively, using standard symmetry fractionalization arguments, we derive $\hat{\mathcal I} \hat R_x^R \hat{\mathcal I}^\dagger = e^{i\theta} \hat R_x^R$ (see Appendix~\ref{app:anom}). Taking these two facts together, we see that a gapped symmetric state implies that $e^{i\theta} =1$, i.e., $\theta = 0$. If $\theta = \pi$, the assumption of a gapped symmetric ground state must be violated.

In the Mott limit, we have $\theta = \pi$, such that we rederive the known LSM anomaly. However, since $\theta$ is a robust quantized invariant, we know that the transition is parametrically stable to finite $U$. The only way $\theta$ can change is (1) by closing the fermion parity gap (such that the parity string does not have long-range order), or (2) by spontaneously breaking the duality symmetry $\mathcal D$, such that the charge $e^{i\theta}$ is no longer well-defined. This particular model opts for the second option, as we see in Fig.~\ref{fig:Fig1} (the white line corresponds to spontaneous dimerization). To confirm this interpretation of the phase diagram, we measure $\theta$ by plotting
\begin{equation}
\mathcal O^\pm := \lim_{|n-m| \to \infty} |\langle \hat{\mathcal S}_m^\pm \hat{\mathcal S}_n^\pm \rangle| \textrm{ with } \hat{\mathcal S}_n^\pm := \prod_{k < n} \hat P_k \left( \hat P_{n} \pm 1 \right).
\end{equation}
It is easy to see that $\mathcal O^\pm \propto |1\pm e^{i\theta}|$, which is nonzero if $e^{i\theta} = \pm 1$. From Fig.~\ref{fig:3}, we conclude that the gapless Mott insulator (MI, $U>U_{c_2}$) has $\theta = \pi$, the bond insulator (BI, $U<U_{c_1}$) has $\theta = 0$, and the intermediate spontaneous dimerized insulator (SDI) has no well-defined $\theta$ since $\mathcal D$ is broken.

\ifx\targetformat\undefined
\textbf{General emergent anomalies.}
\else
\section*{General emergent anomalies}
\fi
The arguments for the parametric stability of edge modes and phase transitions readily extend to other symmetry groups and dimensions. Edge modes of 1D SPT phases are characterized by a non-trivial projective representation of a symmetry group $\tilde G$ \cite{Pollmann10,Turner11,Fidkowski11,Chen11,Schuch11}. One can always extend the symmetry group by $H$ into a bigger symmetry group $G$ (where $\tilde G = G/H$) such that the lifted representation becomes linear, thereby trivializing the SPT phase \cite{Wang2018,Prakash2018,Tachikawa2020,Thorngren2020,Prakash20}. However, the quantum numbers of the additional symmetry group $H$ still label these distinct representations, and the edge mode remains stable until the boundary undergoes a phase transition where it closes the gap to excitations charged under $H$. In the above example (where $\tilde G=SO(3)$, $H=\mathbb Z_2^f$ and $G = SU(2)$), we had to close the fermion parity gap at the boundary to destroy the edge mode.

SPT transitions are stabilized by the mutual anomaly between the protecting symmetry group $\tilde G$ and the duality symmetry at the transition (e.g., the Lieb-Schultz-Mattis anomaly of the Heisenberg chain as discussed above) \cite{Tsui15,Bultinck19,Tsui19}. This anomaly is lifted when we extend $\tilde G$ by $H$, but we propose that there is always an \emph{emergent anomaly} encoded in the symmetry properties of the string orders associated to the symmetry group $H$ (such as the fermion parity string in Eq.~\eqref{eq:Pstring} being odd under $\mathcal D$). It is interesting to observe that since a QSPT transition involves the whole 1D bulk, the emergent anomaly is characterized by a 1D object. In contrast, an edge mode is a zero-dimensional QSPT phenomenon, and correspondingly it parametric stability is encoded in a 0D charge.

Note that the same mechanism of emergent anomalies in 1D SPT transitions applies to the 1D edge modes of 2D SPT phases.
For instance, if one trivializes the $\mathbb Z_2$-SPT in 2D by extending by a second $\mathbb Z_2$ symmetry $\hat P$ into $\mathbb Z_4$, then we expect that the string order of $\hat P$ will have a non-trivial charge under $\mathbb Z_4$---encoding the emergent anomaly. In the simplest case, where the (now non-anomalous) edge mode is written as a 1D system, this string order can be directly computed and its non-trivial charge is a mechanism for the recently-discovered intrinsically gapless SPT phases \cite{Thorngren20b}. In the full 2D model, it can be tricky to directly access this string order (similar to accessing $P_L$ directly in Fig.~\ref{fig:2}), but a clear-cut implication is that the edge mode has an emergent anomaly that can only disappear by spontaneously breaking the symmetry or by driving a boundary phase transition where the gap associated to the extending symmetry $\hat P$ must close. This situation is analogous to what we saw for the SPT transition in the ionic Hubbard chain.

\ifx\targetformat\undefined
\textbf{Discussion.}
\else
\section*{Discussion}
\fi
There are two ways of trivializing an SPT phase without driving through a quantum phase transition: either one \emph{breaks} or \emph{extends} the symmetry group. The former will immediately gap out the edge modes and phase transitions. In this work, we have shown how the latter approach---extending the symmetry group---leaves various topological phenomena intact over a finite region of the phase diagram. We characterized this stability in terms of discrete observables.

This is of particular relevance to physical implementations of SPT phases, where the protecting symmetry group is often a low-energy effective quotient group. For instance, the spin-rotation symmetry protecting the Haldane phase in a spin chain will be derived from a fermion symmetry of an underlying Hubbard model. The results in this work thus show that the experimental realization of such bosonic SPT phases is meaningful, since many of their topological phenomena are stable even when the bulk is---strictly speaking---not in a true SPT phase.

In the above lattice model, we saw a (parametrically) stable transition and edge mode. It is an open question how closely these are linked: it is tempting to think that the emergent edge mode cannot disappear before the emergent phase transition gaps out. Indeed, in Fig.~\ref{fig:Fig1} both features terminate at $U_{c_1}$. This is intuitive given the interpretation of edge modes as being spatially-localized phase transitions \cite{Jackiw76,Teo10,Fukui12,Verresen20}, but it would be interesting to make this correspondence more exact. Another interesting issue is to show that the phase transition is parametrically stable even when the duality symmetry is emergent.

Finally, let us point out that recently the concept of \emph{unnecessary criticality} was introduced, referring to generic continuous phase transitions within the same phase \cite{Bi19,Jian20}. Our work shows that QSPT phenomenology is a general mechanism for constructing such examples in arbitrary dimensions. Intriguingly, 1D examples had not been described before.

\section*{Acknowledgements}
R.V. thanks Ryan Thorngren and Ashvin Vishwanath for discussions in a related collaboration \cite{Thorngren20b}. Calculations were performed using the TeNPy Library \cite{Hauschild18}. R.V. is supported by the Harvard Quantum Initiative Postdoctoral Fellowship in Science and Engineering and by a grant from the Simons Foundation (\#376207, Ashvin Vishwanath). F.P. acknowledges the support of the DFG Research Unit FOR 1807 through grants no. PO 1370/2-1, TRR80, and the Deutsche  Forschungs-gemeinschaft (DFG, German Research Foundation) under Germany's Excellence Strategy EXC-2111-390814868.

\bibliography{main.bbl}

\widetext
\ifx\targetformat\undefined
\begin{center}
	\textbf{\large Supplemental Materials}
\end{center}

\setcounter{equation}{0}
\setcounter{figure}{0}
\setcounter{table}{0}
\makeatletter
\renewcommand{\theequation}{S\arabic{equation}}
\renewcommand{\thefigure}{S\arabic{figure}}
\renewcommand{\bibnumfmt}[1]{[S#1]}
\renewcommand{\citenumfont}[1]{S#1}

\else
\appendix
\fi

\section{Details about symmetry fractionalization}

\subsection{A brief recap of symmetry fractionalization \label{app:symfrac}}

Let us briefly review the argument for symmetry fractionalization \cite{Turner11}, focusing on 1D systems for concreteness.

If $|\psi\rangle$ is the ground state of a gapped local quantum Hamiltonian, it will have a finite correlation length $\xi$ \cite{Hastings06}. For any block of $N$ sites, we can consider the Schmidt decomposition $|\psi\rangle = \sum_\alpha \Lambda_\alpha |\psi^\textrm{in}_\alpha\rangle \otimes |\psi^\textrm{out}_\alpha\rangle$, where $|\psi^\textrm{in}_\alpha\rangle$ are quantum states defined purely within the block of $N$ sites. Due to there being a finite correlation length, all the Schmidt states $|\psi^\textrm{in}_\alpha\rangle$ are indistinguishable deep within the block (i.e., they simply look like $|\psi\rangle$). Moreover, due to locality, the two edges of the block are effectively independent (up to exponentially small errors). We can thus presume that $\alpha$ can be interpreted as a super-index which is equivalent to two smaller indices $\alpha_L,\alpha_R$ such that $|\psi^\textrm{in}_\alpha\rangle = |\psi^\textrm{in}_{\alpha_L,\alpha_R}\rangle $ where the value of $\alpha_L$ ($\alpha_R$) only affects correlation functions on the left (right) end of the block.

Let $\hat U = \prod_n \hat U_n$ be a unitary on-site symmetry. If we act with $\hat U$ only within this block, it leaves the states $|\psi^\textrm{out}_\alpha\rangle$ unaffected. Since the states in a Schmidt decomposition form a complete basis, we can write $\hat U |\psi^\textrm{in}_\alpha\rangle = \sum_\beta U_{\beta;\alpha} |\psi^\textrm{in}_\beta\rangle = \sum_{\beta_L, \beta_R} U_{\beta_L, \beta_R; \alpha_L, \alpha_R} |\psi^\textrm{in}_{\beta_L,\beta_R}\rangle$. Since $\hat U$ is an on-site operator, it will preserve locality, i.e., $\alpha_L$ and $\beta_R$ in $U_{\beta_L, \beta_R; \alpha_L, \alpha_R}$ will be uncorrelated. More precisely, we can write $U_{\beta_L, \beta_R; \alpha_L, \alpha_R} = U_{\beta_L; \alpha_L}^L U_{\beta_R; \alpha_R}^R$. In short, we write that $\prod_{n=1}^L \hat U_n |\psi\rangle = \hat U^L \hat U^R |\psi\rangle$ where $\hat U^{L,R}$ are exponentially localized near the edges of the block of $N$ sites. The explicit formula in terms of Schmidt states implies that if $\prod_{i=1}^L \hat U_i |\psi\rangle = \hat U^L \hat U^R |\psi\rangle$ and $\prod_{i=1}^L \hat V_i |\psi\rangle = \hat V^L \hat V^R |\psi\rangle$ (for a second symmetry $\hat V$), then we also have that $\prod_{i=1}^L \hat V_i \prod_{i=1}^L \hat U_i |\psi\rangle = \hat V^L \hat V^R \hat U^L \hat U^R |\psi\rangle$. (This is not as obvious as it might seem on first sight, since the conclusion would not hold if the $\hat U$ and $\hat V$ symmetries were acting on distinct-but-overlapping blocks).

The fractionalized symmetries obey the same group relations as the original symmetries up to potential phase factors. For instance, let us suppose $\hat U$ and $\hat V$ commute. Let us also assume that $\hat U^{L,R}$ are bosonic operators (such that $\hat U^L$ commutes with $\hat V^R$). Then
\begin{equation}
1 = \hat U \hat V \hat U^{-1} \hat V^{-1} = \hat U^L \hat V^L \big( \hat U^L\big)^{-1}  \big( \hat V^L\big)^{-1} \times \hat U^R \hat V^R \big( \hat U^R\big)^{-1}  \big( \hat V^R\big)^{-1}.
\end{equation}
Since the two factors on the right-hand side act on disjoint regions yet they multiply to the identity, each of the two factors has to be proportional to a number: $\hat U^L \hat V^L \big( \hat U^L\big)^{-1}  \big( \hat V^L\big)^{-1} = e^{i\alpha}$. Moreover, using similar manipulations, one can show that if $\hat U^2= 1$, then $e^{i\alpha} = \pm 1$. More generally, the fractionalized symmetries will form a projective representation of the original symmetry group. Non-trivial projective representations correspond to non-trivial SPT phases and imply edge modes in the energy spectrum with open boundary conditions, or degeneracies in the entanglement spectrum for virtual bipartitions (since any projective representation acting on a 1D vector space is trivial).

\subsection{Derivation of the $\theta$ invariant \label{app:theta}}

It is instructive to derive the invariant $\theta$ in Eq.~\eqref{eq:Pstring} rigorously, which we can do using symmetry fractionalization. For a parity string which is much longer than the fermionic correlation length, we can write $\hat{\mathcal P}_{m,n} := \prod_{m\leq k \leq n} \hat P_k = \hat P^L_m \hat P^R_n$ on the ground state subspace. Since $\hat{\mathcal D} \hat{\mathcal P}_{m,n} \hat{\mathcal D}^\dagger = \hat{\mathcal P}_{m+1,n+1}$, we obtain
\begin{equation}
\left( \hat{\mathcal D} \hat P_m^L \hat{\mathcal D}^\dagger \right) \left( \hat{\mathcal D} \hat P_n^R \hat{\mathcal D}^\dagger \right) =\hat P^L_{m+1} \hat P^R_{n+1} \quad \Rightarrow \quad \hat{\mathcal D} \hat P_m^L \hat{\mathcal D}^\dagger = \alpha_{m,n} \hat P_{m+1}^L \textrm{ and } \hat{\mathcal D} \hat P_n^R \hat{\mathcal D}^\dagger = \bar \alpha_{m,n} \hat P_{n+1}^R,
\end{equation}
where $\alpha_{m,n}$ is some proportionality factor. Note that the first (second) equation tells us that it cannot depend on $n$ ($m$). I.e., the proportionality factor is a genuine constant. Let us denote it as $\alpha_{n,m} =e^{i\theta}$. Since $\langle \hat{ \mathcal P}_{m,n} \rangle = \langle \hat P^L_m \rangle \langle \hat P^R_n \rangle$ (due to locality and the spatial separation between the fractionalized symmetries $\hat P^{L,R}$) and the fact that $\hat{\mathcal D}$ is a symmetry (i.e., $\langle \hat P^L_m \rangle = \langle \hat{\mathcal D} \hat P^L_m \hat{\mathcal D}^\dagger \rangle = e^{i\theta} \langle \hat P^L_{m+1} \rangle = e^{i\theta k} \langle \hat P^L_{m+k} \rangle$), we derive Eq.~\eqref{eq:Pstring}:
\begin{equation}
\langle \hat P_{m} \hat P_{m+1} \cdots \hat P_{n-1} \hat P_n \rangle 
= \langle \hat{ \mathcal P}_{m,n} \rangle = \langle \hat P^L_m \rangle \langle \hat P^R_n \rangle
= e^{i\theta(n-m)} \langle \hat P^L_{n_0} \rangle \langle \hat P^R_{n_0} \rangle,
\end{equation}
where $n_0$ is some reference site that does not depend on $n$ or $m$. Finally, the quantization of $\theta$ follows from $\hat{\mathcal P}_{m,n}^2 = 1$ since then $( \hat P^R_n )^2 \propto 1$ such that $e^{2i\theta} = 1$, i.e., $\theta=0,\pi$. (As an illustration, note that in the spin chain limit, $P^{R}_n = (-1)^n$, which implies $\theta = \pi$.)

\subsection{The emergent anomaly \label{app:anom}}

Here we derive that $\hat{\mathcal I} \hat R_x^R \hat{\mathcal I}^\dagger = e^{i\theta} \hat R_x^R$ where $\hat{\mathcal I} = \hat R_y \hat{\mathcal D} \hat R_y^\dagger \hat{\mathcal D}^\dagger$. This straightforwardly follows from symmetry fractionalization. We will need what we derived above: $\hat{\mathcal D} \hat P_n^L \hat{\mathcal D}^\dagger = e^{i\theta} \hat P_{n+1}^L$. Similary, there is a phase factor $e^{i\kappa}$ such that $\hat{\mathcal D}\big[ \hat R_x^L\big]_n \hat{\mathcal D}^\dagger = e^{i\kappa} \big[ \hat R_x^L\big]_{n+1}$. Lastly, since group relations are always obeyed up to a phase factor, there is phase factor $e^{i\mu}$ such that
\begin{equation}
\hat R_y^\dagger  \big[\hat R_x^L\big]_{n}  \hat R_y = \big[ \hat R_y^L \big]_n^\dagger \big[\hat R_x^L\big]_{n} \big[\hat R_y^L\big]_{n} = e^{i\mu} \hat P^L_n \big[\hat R_x^L\big]_{n} .
\end{equation}
Note that this also implies $\hat R_y \hat P^L_n \big[\hat R_x^L\big]_{n}  \hat R_y^\dagger = e^{-i\mu}\big[\hat R_x^L\big]_{n}  $. Plugging in these identities, we obtain
\begin{align}
\boxed{\hat{\mathcal I} \big[\hat R_x^L\big]_n \hat{\mathcal I}^\dagger} &= \hat R_y \hat{\mathcal D} \hat R_y^\dagger \big( \hat{\mathcal D}^\dagger  \big[\hat R_x^L\big]_n \hat{\mathcal D} \big) \hat R_y \hat{\mathcal D}^\dagger R_y^\dagger = e^{-i\kappa} \hat R_y \hat{\mathcal D}\big(\hat R_y^\dagger  \big[\hat R_x^L\big]_{n-1}  \hat R_y \big) \hat{\mathcal D}^\dagger R_y^\dagger  \\
&= e^{i(\mu-\kappa)} \hat R_y \hat{\mathcal D}  \hat P_{n-1}^L\big[\hat R_x^L\big]_{n-1}  \hat{\mathcal D}^\dagger R_y^\dagger
= e^{i(\mu-\kappa)} \hat R_y \big( \hat{\mathcal D}  \hat P_{n-1}^L \hat{\mathcal D}^\dagger \big) \big( \hat{\mathcal D}\big[\hat R_x^L\big]_{n-1}  \hat{\mathcal D}^\dagger \big) R_y^\dagger \\
&= e^{i(\mu-\kappa)} \hat R_y \big( e^{i\theta}  \hat P_{n}^L  \big) \big( e^{i\kappa} \big[\hat R_x^L\big]_{n} \big) R_y^\dagger = e^{i\theta} \big(e^{i\mu} \hat R_y  \hat P_{n}^L  \big[\hat R_x^L\big]_{n}  \hat R_y^\dagger \big) = \boxed{ e^{i\theta } \big[ \hat R_x^L\big]_{n} } .
\end{align}
\end{document}